# Explanatory Analysis and Rectification of the Pitfalls in COVID-19 Datasets

Samyak Prajapati*
Department of Computer Science and Engineering
National Institute of Technology Delhi
New Delhi, Delhi, India
181210046@nitdelhi.ac.in

Japman Singh Monga*
Department of Computer Science and Engineering
National Institute of Technology Delhi
New Delhi, Delhi, India
181210024@nitdelhi.ac.in

Shaanya Singh*
Department of Computer Science and Engineering
Pandit Deendayal Energy University
Gandhinagar, Gujarat, India
shaanya.sce18@sot.pdpu.ac.in

Amrit Raj*
Department of Computer Science and Engineering
National Institute of Technology Delhi
New Delhi, Delhi, India
181210008@nitdelhi.ac.in

Yuvraj Singh Champawat*
Department of Computer Science and Engineering
National Institute of Technology Delhi
New Delhi, Delhi, India
181210064@nitdelhi.ac.in

Dr Chandra Prakash
Department of Computer Science and Engineering
National Institute of Technology Delhi
New Delhi, Delhi, India
cprakash@nitdelhi.ac.in

*Co-First Authors

**ABSTRACT**

Since the onset of the COVID-19 pandemic in 2020, millions of people have succumbed to this deadly virus. Many attempts have been made to devise an automated method of testing that could detect the virus. Various researchers around the globe have proposed deep learning based methodologies to detect the COVID-19 using Chest X-Rays. However, questions have been raised on the presence of bias in the publicly available Chest X-Ray datasets which have been used by the majority of the researchers. In this paper, we propose a 2 staged methodology to address this topical issue. Two experiments have been conducted as a part of stage 1 of the methodology to exhibit the presence of bias in the datasets. Subsequently, an image segmentation, super-resolution and CNN based pipeline along with different image augmentation techniques have been proposed in stage 2 of the methodology to reduce the effect of bias. InceptionResNetV2 trained on Chest X-Ray images that were augmented with Histogram Equalization followed by Gamma Correction when passed through the pipeline proposed in stage 2, yielded a top accuracy of 90.47% for 3-class (Normal, Pneumonia, and COVID-19) classification task.

## 1. INTRODUCTION

During the late months of 2019, a series of pneumonia cases were detected in the Hubei province in mainland China, and soon after, the disease was named COVID-19 by the World Health Organization. The contagion was identified to be of the SARS (Severe Acute Respiratory Syndrome) species and formally termed as SARS-CoV-2, exposure to which engenders multiple respiratory illnesses in humans. It is a highly contagious respiratory disease that is spread by direct or indirect contact from infected bodies, and this is reasoned to be the prime driver of its high transmissibility. This high transmissibility of the virus is corroborated by the fact that the $R_0$ value of the ancestral strain was estimated to be between 2-3, whereas the current sweeping Delta variant (B.1.617.2) is estimated to have an $R_0$ of 5-9 [1]. The virion has infected over 209 million individuals and is responsible for the deaths of over 4 million individuals, forcing countless others to isolate themselves in fear of their safety [2]. In the absence of quintessential treatment procedures, the contagious SARS-CoV-2 has caused a surge in infection rates around the world (Figure 1) and has ruptured the concept of normal life.

A notion of vaccine hesitancy among certain sects of society may also lead to an influx of variants of concern (VOCs) that may be the product of convergent evolution, thus gaining traits like higher transmissibility and resistance to antibody action [3]. In vitro experiments have shown that the currently rampant Delta variant (B.1.617.2) is 8 folds less sensitive to vaccine-elicited antibodies when compared to the wild type (Wuhan 1) [4], thus inflicting multiple "waves" of COVID cases across the globe.

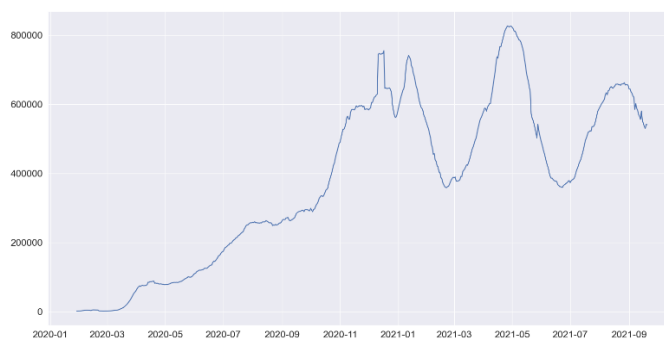

*Figure 1: Variation of the magnitude of daily occurrences of COVID-19 cases through the pandemic* [2]

As evident with the "spikes" of cases of COVID-19, rapid deployment of on-ground personnel and rapid testing is crucial in containing these unpredictable surges. Accurate predictions of these surges and swift upscaling of the healthcare framework are critical towards ensuring the availability of appropriate facilities during these unfortunate conditions of a pandemic [5]. Some countries may exhibit the need to divert their manpower into the assembly of physical infrastructure, whereas others may be in the need of securing the capital for further investment in the aforementioned infrastructure. In such cases, automated methods of diagnosis would be necessary for reducing the load on the healthcare infrastructure and increasing the efficiency of the detection of COVID-19 in individuals [6]. While there have been efforts in devising an automated approach for such a case, they usually employ the use of the publicly available datasets, which are prone to the introduction of unintended biases when left unchecked, and in such a case, we propose a new methodology that is aimed at eliminating such biases. The rest of this paper is organized as follows; section 2 describes the literature review of similar works, section 3 provides insights about the raw data that was used in this study, with section 4 describing the methodology used. section 5 has details pertaining to the performance metrics used to evaluate the models and section 6 and 7 contain the results and the conclusions of this study respectively.

## 2. LITERATURE REVIEW

In this section, we describe some of the exceptional topical research work, motivation and salient factors that have governed the proceedings of this study.

Studies from around the globe have focused on developing a computer vision based methodology which could help in screening and classifying COVID-19 cases. Wang et al. in their work [7] proposed a tailored transfer learning based CNN architecture 'Covid-Net' to perform a 3-class classification (Pneumonia, Normal and COVID-19), achieving 93.3% Top-1 accuracy, which was comparatively better than a simple transfer-learning based approach. Arpan Mangal et al. went a notch further and proposed a 4-class classification (COVID-19, Normal, Viral and Bacterial Pneumonia) methodology along with 3-class classification [8]. They obtained 87.2% accuracy and 90.5% accuracy for 4-class and 3-class classification respectively. It is important to note that the aforementioned studies were conducted on datasets formed by taking a combination of images from multiple open-source datasets.

Many studies which have combined multiple open-source datasets to develop a deep learning based architecture for the classification of COVID-19 have been published. However, M.G and N.L's work [9] was one of the first studies to point out a shortcoming of these approaches. In their work, they augmented 4 publicly available datasets by using different operations and subsequently placed a black mask over the images to block the center of the image, obscuring the lung from the images. The images were trained on AlexNet to classify COVID and non-COVID CXR images. Their methodology was able to achieve a minimum AUC-ROC score of 0.92 and scoring as high as 0.999 with obscured lung regions. This corroborated that there was indeed a bias in the datasets and that the model was in fact learning distinguishing features between the datasets instead of distinguishing the presence of COVID-19. Further research in the field proved this point. Robert et al. [10] published a review work, in which the risk of bias was assessed for 62 papers using the PROBAST tool defined in the study. The findings of this review work proved that the majority of the papers had shortcomings such as bias in the datasets, and poor reproducibility. Santa Cruz et al in their work [11], used PROBAST and CHARMS tools to assess the high risk of bias in medical imaging due to the absence of information which is vital to assess selection bias, and lack of clarity of the procedures of image augmentation.

Therefore, it is crucial that we select our datasets carefully and apply data augmentation & preprocessing methods to reduce the bias in the datasets. Chowdhury et al. [12] conducted 2 and 3 class classification with models trained on images with and without augmentation. DenseNet201 outperformed the others with training being done on images with augmentation. R. Kushol et al. in their work [13] have proposed an image enhancement technique that consists of applying combined Top-hat and Bottom-hat transform and finding Optimal Structuring Element (SE) size and subsequently comparing their method to established image enhancement methods such as CLAHE. Their proposed method generates comparatively clear and visually improved output. Hence, in this study, we've established the presence of bias in the publicly available datasets, and subsequently proposed a transfer learning based pipeline with image processing and augmentation techniques to diminish the bias.

## 3. DATA SOURCE

Multiple publicly available datasets were analysed and selected with careful consideration to eliminate the formation and the usage of "Frankenstein" datasets [10][11]. The datasets used in the study are:

- **Chest X-Ray Images (Pneumonia) Dataset** [14]
  This dataset was published by Paul Mooney on Kaggle and is commonly referred to as the "Kaggle Dataset", "Paul Mooney Dataset" or the "Kermany Dataset". It consists of 5856 pediatric images that were classified into "Normal", "Bacterial Pneumonia" and "Viral Pneumonia" from the Guangzhou Women and Children's Medical Center. This dataset was one of the major open-source datasets available for Viral and Bacterial Pneumonia and consequently, it has been used in numerous studies.

- **CheXpert Dataset** [15]
  This dataset contains 224,316 chest radiographs images which were collected from Stanford Hospital between October 2002 and July 2017 across 65,240 patients. The collected CXR images were further analyzed and labelled for the presence of 14 biological observations. It consists of a total of 16,627 Normal images (with no indication of any diseases) and 4,576 images with the presence of Pneumonia.

- **RSNA Dataset** [16]
  This dataset was created by the Radiological Society of North America (RSNA) in collaboration with the US National Institutes of Health (NIH). It consists of 26,684 Chest X-Ray (CXR) imagery in DICOM format, split between pneumonia (9,555 images) and non-pneumonia (20,672 images).

- **BrixIA COVID-19 Dataset** [17]
  This dataset was collected at the ASST Spedali Civili di Brescia, Italy during their first peak of COVID cases (March 4 - April 4, 2020). It consists of 4,707 CXR images of patients with confirmed cases of COVID-19.

- **Montgomery County Chest X-Ray Dataset** [18]
  This dataset contains 138 images split across "Normal" (80 images) and "Manifestation of Tuberculosis" (58 images) from the Department of Health and Human Services, Montgomery County, Maryland, USA. Lung segmentation masks were also made available in the original dataset.

- **Shenzhen Chest X-Ray Dataset** [18]
  This dataset contains 662 CXR images from the Shenzhen No.3 People's Hospital, Guangdong Medical College, Shenzhen, China. It was split across "Normal" (326) and "Manifestation of Tuberculosis" (336). Lung segmentation masks were not available in the original dataset, however, the works of S. Candemir et al. have published their manual lung segmentation masks on Kaggle [19]. It is important to note that this dataset contains pediatric CXR images as well.

## 4. METHODOLOGY

In this paper, a two-stage methodology has been proposed, where stage 1 establishes the presence of bias in publicly available COVID-19 datasets and thereafter, a novel pipeline is proposed in stage 2 to reduce the presence of bias.

### 4.1. Stage 1

Our main concern in this section was to confirm the presence of bias in the popular datasets that were used in the classification of COVID-19 from the types of pneumonia in CXR imagery [9]. In order to do so, two separate experiments were conducted as a part of stage 1 of our work.

#### 4.1.1. t-SNE + SVM

t-Distributed Stochastic Neighbor Embedding or t-SNE is an unsupervised, non-linear technique that is primarily used for dimensionality reduction, data exploration and visualizing high-dimensional data. Five datasets, CheXpert (Normal and Pneumonia), RSNA (Normal and Pneumonia), Paul Mooney (Normal and

Pneumonia) Montgomery (Normal and Others) and Shenzhen (Normal and Others) datasets were considered for evaluation in this section. 200 Chest X-ray images from each class of CheXpert, RSNA, Paul Mooney, Shenzhen datasets and 80 images of 'Normal' class, and 58 images of 'Others' class from Montgomery dataset were sampled randomly to create 'Mini-Dataset A'.

Different combinations of the same labelled classes- for instance, the 'Normal' class of CheXpert and the 'Normal' class of RSNA- were created from the Mini-Dataset A. Images were resized to 256x256 pixels, followed by normalization of the pixel data to a value between 0 and 1. For each combination, a train-test split of 90%-10% was chosen for evaluation.

t-SNE algorithm with 2 components on pixel intensity was applied on each combination. Subsequently, an SVM classifier was trained on the 2-D features extracted from the t-SNE algorithm to perform the binary classification between the same labelled classes. The trained SVM classifier was used to evaluate the test accuracy for each combination. Our prime motivation to perform this experiment was to prove the bias in the datasets. In such a classification task, i.e. in a classification task between the same labelled classes, a high test accuracy implies that the trained SVM classifier can detect and classify the source dataset of the image, hence proving that bias exists in the datasets.

Thereafter, eight different preprocessing techniques were applied to the Mini-Dataset A and the aforementioned experiments was repeated with each of the preprocessing techniques. Subsequently, test accuracies were compared with the initial ones and the shift in the accuracies was noted.

### 4.1.2. Grad-CAM

Grad-CAM is a technique under the umbrella of Explainable-AI (XAI) that is used to generate "visual explanations" for the decisions made in CNNs [20]. Three datasets, Paul Mooney, CheXpert and the RSNA datasets were considered in this section for evaluation. All these datasets have two classification labels, "Normal" and "Pneumonia". In order to generate a comparison between the datasets, 500 images from each class were randomly sampled from each of the datasets to create a "Mini-Dataset B" for training a model and a 90-10 split was chosen for train-validation evaluation. In this case, a feed-forward model (VGG-16) was chosen without its FC layers, instead, the outputs were flattened and a dropout layer was applied, which was followed by a 128-unit Dense layer, and a 2-unit softmax activation layer as illustrated in Figure 2.

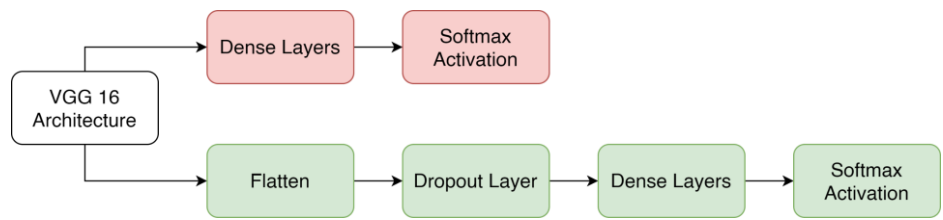

*Figure 2: Comparison of the original architecture (red) with the chosen architecture (green)*

The images were augmented by performing random rotation, random horizontal flipping and resized to 36x36 pixels, followed by normalization of the pixel data to a value between 0 and 1. The prime motivation for the drastic reduction of image resolution was to highlight the absence of spatial data, which would be crucial in determining the presence of COVID-19. A similar test set was also created with 500 images from each classification class. This data was then fed into the model for training and testing on multiple combinations of the three datasets. An unconventional train-test split was used to demonstrate the fact, that the model was able to learn to classify the images despite the meagre size of the training dataset.

### 4.2. Stage 2 - Proposed Methodology

The primary goal of stage 2 was to generate a comparison between different augmentation techniques and their effects on reducing the inherent bias between different datasets. In order to do so, a data pipeline was proposed (Figure 3) that uses the concepts of supervised learning with image segmentation, image super-resolution and image augmentation. A combined total of 52,594 images from the CheXpert, RNSA and BrixIA datasets were fed into the pipeline and saved locally.

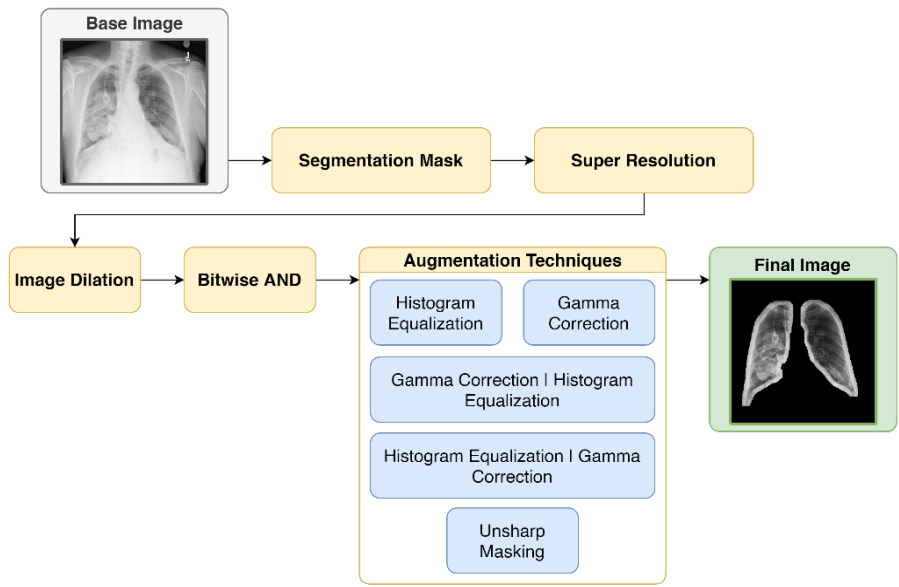

*Figure 3: Proposed pipeline of stage 2*

### 4.2.1. Image Segmentation

U-Net is a popular image segmentation model developed by Ronneberger et al. [21] that is widely used in the biomedical domain due to its ability to outperform traditional sliding-window based networks in the task of segmentation. The model architecture exhibits a symmetrical encoder-decoder based model. The "skip-connections" from the contracting (encoder) section allows the model to combine features from different spatial regions of the image to the expanding (decoder) section, thus granting it excellent capacity in developing knowledge about the features and their spatial significance. In our work, U-Net was used to create segmentation masks of cheXpert, RSNA and BrixIA datasets. The images were first sized to a size of 256x256 in order to minimize the development of certain artifacts as illustrated in Figure 4, the resized images were then fed into the U-Net model for generating the segmentation masks. The segmented lungs served the purpose of eliminating the extraneous region that was considered irrelevant in detecting the presence of a pulmonary malady.

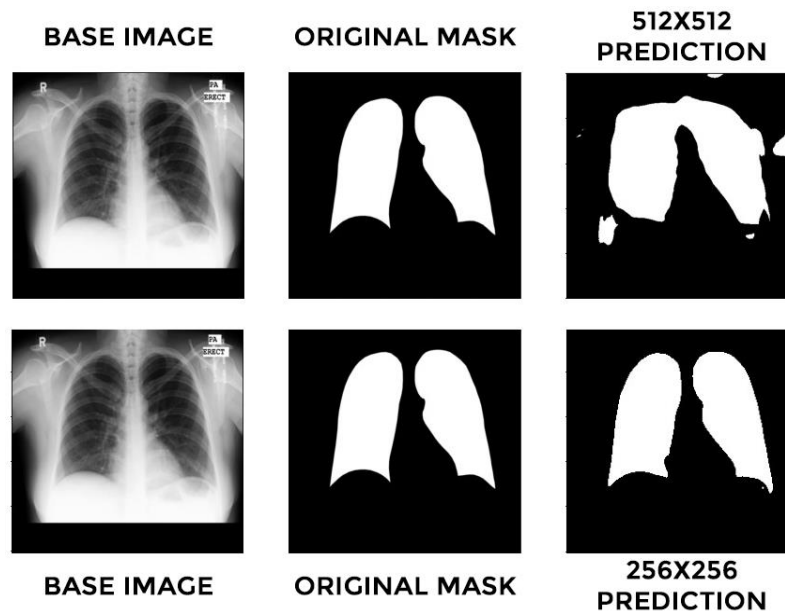

*Figure 4: Comparison of generated segmentation mask of input image sizes of 512x512 (top row) and 256x256 (bottom row)*

### 4.2.2. Super-Resolution

Super-resolution is an image upscaling technique that is commonly used in the medical domain to generate high-resolution images from certain images with lower than satisfactory resolution. ESRGAN [22] makes use of the general philosophy of generative adversarial networks (GANs) and builds upon it to increase their performance. The works of Wang et al. makes three key changes to the workings of the SRGAN [23],

1. They make use of the Residual-in-Residual Dense Block (RDDB) and eliminate the use of Batch Normalization layers. The combination of residual scaling and smaller initializations allows them to train a very deep network.
2. They improve their discriminator by the use of the Relativistic average GAN (RaGAN) [24], which learns to judge images on the basis of their authenticity. This allows the generator to recover realistic texture details.
3. They propose the use of perceptual loss, which utilizes VGG features before activation instead of after activation as in SRGAN. Empirical evidence suggests that the use of perceptual loss presents sharper edges and visually pleasing results.

In our work, the 256x256 image segmentation mask was upscaled using ERSGAN and then resized to the original image size. The binary segmentation masks were dilated so as to smoothen any jagged edges and then merged with the original image using a "bitwise and" operation.

### 4.2.3. Image Augmentation

Image augmentation techniques have the potential to scramble possible confounding data (to an extent), and due to this very nature, they can be used, along with other techniques to minimize the presence of bias in our datasets. In order to accomplish this goal, five combinations of the following augmentation techniques have been implemented and tested.

#### 4.2.3.1. Unsharp Masking

Unsharp Masking is a method of linear image processing that helps make the images sharper [25]. This helps the model to notice the important features of the image. A mask is created by obtaining a blurred version of the image and taking the value of its difference with the original image by scaling it. This amount is then added back to the original to obtain the final sharpened image. The blurred version of the image can be created using different imaging filters. In our methodology, we have used the gaussian filter. The formula of obtaining the enhanced image can mathematically be represented as:

$$Final\ Image = Original\ Image + Amount \times (Original\ Image - Blurred\ Image)$$

The blurring of the image is controlled by the Radius parameter that is passed through the unsharp masking function. The radius amount is basically the sigma value in the gaussian filter.

#### 4.2.3.2. Histogram Equalization

Histogram Equalization is a method of image processing that helps bring contrast in images. Histogram Equalization cannot be applied to separate colour components such as Red, Green or Blue on their own, as it can lead to an imbalance in colour concentration. Hence, the image has to be converted into another colour space (HSL/HSV) following which the method can be applied without causing changes in the saturation of the image. This method works by spreading out the pixel intensity values, thus allowing areas with low contrast to gain more contrast. The method of creating a uniform distribution of pixels is a monotonic and non-linear method which results in a flat histogram. It helps in detail enhancement and increases the contrast of the image globally.

#### 4.2.3.3. Gamma Correction

Gamma correction is a non-linear method used to change the brightness of an image. To apply the gamma correction, the pixel intensities have to be scaled from [0,255] to [0,1.0]. The equation of the gamma correction function is as follows:

$$O = I^{(1/G)}$$

Here G is the gamma value, O is the output which is scaled back to [0,255] and I is the input image. The value of G determines if the image will shift towards the darker side or the lighter side. If the value of G is less than 1 then the image will appear darker whereas if the value is greater than 1 then the image will appear lighter. If the value of G is equal to 1 then there will be no change in the image.

For our study, we've employed 5 combinations of the aforementioned preprocessing techniques; histogram equalization, gamma correction, histogram equalization followed by gamma correction, gamma correction followed by histogram equalization, and unsharp masking. Gamma value of 1.5 was used for each of the relevant techniques.

#### 4.2.4. Training Model

In order to stay on track with the aim of stage 2, two convolutional neural network (CNN) based models were trained and evaluated on different augmentation techniques. The 52,594 locally saved images were split into their training and testing splits with a 90:10 ratio, and the class imbalance was tackled by assigning weights to each class. An empirical evaluation was done on DenseNet-101 and InceptionResNetV2 to quantitatively compare the effects of bias reduction by augmentation techniques.

## 5. RESULTS

### 5.1. Performance Metrics

Since the experiment was conducted in two stages, stage 1 solely used Top-1 Accuracy as its quantitative metric, whereas, Top-1 Accuracy, Precision, Recall, AUROC and F1 Scores were the metrics used in stage 2 for establishing the performance of the models. These metrics can be broken down into their components - True Positives (TP), False Positives (FP), False Negatives (FN) and True Negatives (TN) using the following formulae.

$$Accuracy = \frac{TP + TN}{TP + TN + FP + FN}$$

$$Precision = \frac{TP}{TP + FP}$$

$$Recall = \frac{TP}{TP + FN}$$

$$F1\ Score = 2 * \frac{Precision * Recall}{Precision + Recall}$$

It is important to note that the AUROC (Area under ROC) is a metric that allows us to attain a degree of separability, where a value of 1 would indicate that there is a perfect separation between the different classes and a value of 0.5 would be an indication of a model that is not able to generate a prediction at all, and hence, is making random guesses.

### 5.2. Stage 1

In this section, we present the results of the first stage of our experiment. This step was performed to confirm the bias present in the datasets used for the classification of COVID-19. As stipulated in the previous section, two experiments were conducted as a part of stage 1, the results of which are given below.

#### 5.2.1. t-SNE + SVM

As stated in the methodology section, this experiment was performed on different combinations of the same labelled classes from the Mini-Dataset A. t-SNE was applied on each combination- for instance, 'Normal' class of RSNA and 'Normal' class of CheXpert- and 2-D features were extracted which were subsequently fed to an SVM classifier for the classification task. The results are stipulated in table 1 below.

Table 1: Accuracy obtained with SVM on different dataset combinations with 'N' representing Normal and 'P' representing Pneumonia

| | Dataset Combination | | Accuracy |
|---|---|---|---|
| (1) | RSNA (N) | CheXpert (N) | 0.9 |
| (2) | Paul Mooney (N) | RSNA (N) | 0.975 |
| (3) | Paul Mooney (N) | CheXpert (N) | 1 |
| (4) | RSNA (P) | Paul Mooney (P) | 0.975 |
| (5) | CheXpert (P) | RSNA (P) | 0.925 |
| (6) | CheXpert (P) | Paul Mooney (P) | 1 |

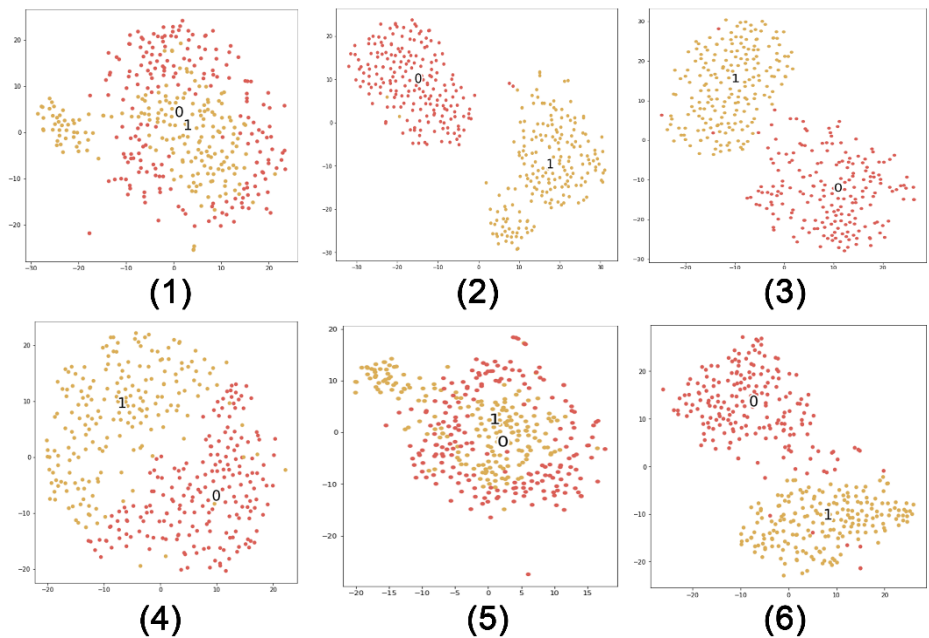

Figure 5: t-SNE plots of dataset combinations as described in Table 1

As evident from the t-SNE plots of Figure 5 and the results of Table 1, the usage of the Paul Mooney dataset leads to a clear separation of data points, thus introducing an unnatural increase in the accuracy, this is now attributed to the presence of extreme bias in this dataset. Due to this, the Paul Mooney dataset was eliminated in our study. Thereafter, the same experiment was repeated with the remaining datasets, but this time, certain image augmentation techniques were applied to draw a comparison in their effectiveness as reducing bias (Table 2). Keeping these metrics in mind, we can come to a conclusion that Histogram Equalization, Histogram followed by Gamma, Gamma followed by Histogram, Gamma 0.5 and Unsharp Masking were effective at bias reduction to an extent, and hence these were the techniques chosen for augmenting the images in stage 2. Also, it is evident that even after the application of augmentation techniques, presence of either Shenzhen or Montgomery datasets leads to an unnaturally high accuracy, thus suggesting a high bias. Hence, both of these datasets were excluded from the further analysis in stage 2.

*Table 2: Accuracy obtained with SVM on different dataset combinations after augmentation with 'N' representing Normal, 'P' representing Pneumonia and 'O' representing Others*

| Dataset Combination | | Augmentation Techniques | | | | | | | |
|---|---|---|---|---|---|---|---|---|---|
| | | Histogram Equalization | Histogram Equalization followed by Gamma Correction | Gamma Correction followed by Histogram Equalization | Gamma Correction (0.5) | Gamma Correction (1.5) | Gamma Correction (2.0) | CLAHE | Unsharp Masking |
| RSNA (N) | CheXpert (N) | 0.65 | 0.6 | 0.675 | 0.775 | 0.725 | 0.775 | 0.8 | 0.775 |
| RSNA (P) | CheXpert (P) | 0.7 | 0.75 | 0.725 | 0.975 | 0.9 | 0.975 | 0.9 | 0.925 |
| Montgomery (N) | Shenzhen (N) | 1 | 1 | 1 | 1 | 1 | 1 | 1 | 1 |
| Montgomery (O) | Shenzhen (O) | 0.923 | 0.962 | 1 | 1 | 1 | 1 | 1 | 1 |
| CheXpert (N) | Montgomery (N) | 1 | 1 | 1 | 0.964 | 0.964 | 1 | 0.964 | 0.964 |
| CheXpert (N) | Shenzhen (N) | 0.975 | 0.975 | 0.975 | 1 | 1 | 1 | 1 | 1 |
| RSNA (N) | Montgomery (N) | 1 | 0.964 | 0.964 | 1 | 1 | 1 | 1 | 1 |
| RSNA (N) | Shenzhen (N) | 0.975 | 0.975 | 0.975 | 0.975 | 0.925 | 0.975 | 1 | 0.975 |

### 5.2.2. Grad-CAM

This experiment was implemented on a combination of Normal and Pneumonia classes from the three datasets (Paul Mooney, CheXpert and RSNA dataset). 'Mini-Dataset B' was created and trained on our VGG-16 based network. The models were trained and tested to evaluate their performance in classifying different classes (Normal & Pneumonia) of multiple datasets, for example, the model was trained on the training subset of RSNA (Normal) and Paul Mooney (Pneumonia) and then tested on the testing subset of the same combination. The results of the experiments are tabulated in Table 3 below.

*Table 3: Accuracy obtained with VGG-16 on different dataset combinations with 'N' representing Normal and 'P' representing Pneumonia*

| Dataset Combination | | Accuracy |
|---|---|---|
| Normal | Pneumonia | |
| RSNA(N) | Paul Mooney(P) | 0.915 |
| RSNA (N) | CheXpert(P) | 0.939 |
| CheXpert(N) | RSNA(P) | 0.98 |
| CheXpert(N) | Paul Mooney(P) | 0.98 |
| Paul Mooney(N) | RSNA(P) | 0.99 |
| Paul Mooney(N) | CheXpert(P) | 0.981 |

As shown in Table 3, the high accuracies obtained for all the different dataset combinations indicates that the model was able to classify the images as Normal or Pneumonia despite a small image size of 36x36 pixels. In such a scenario, the image is void of significant spatial data, which would allow for efficacious classification, thus, it is evident that the model has learnt to classify the images on a completely different feature, thereby suggesting the presence of significant bias in the datasets.

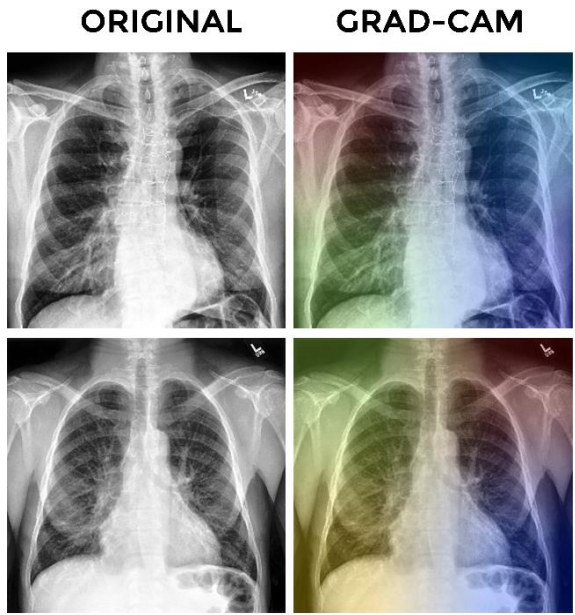

*Figure 6: Original CXR image with corresponding Grad-CAM heat maps*

To further confirm our hypothesis, we made use of Grad-CAM based activation masks, which allows us to visualize the areas of importance that a model chooses in order to make a prediction. Figure 6 is an illustration depicting the same, where we can notice that the superimposed heatmap of the 36x36 image has a greater concentration of neuron activations along the corner when compared with the relative absence of neuron activations in the lung region. It is evident that the model is making predictions with a greater weight to the corner, suggesting the presence of certain information that can be used to differentiate between datasets i.e., a certain type of bias.

### 5.3. Stage 2

In this section, we present the results of the second stage of our experiment, where we trained and tested CNN based models, DenseNet-201 and InceptionResNetV2, on images of individual augmentation techniques (following the pipeline of Figure 3) as selected in stage 1 of our experiment. The results for each model on isolated augmentation techniques are tabulated below in Tables 4 and 5.

*Table 4: Classification metrics of DenseNet-201 on different augmentation techniques*

| Augmentation Technique | Accuracy | Precision | Recall | AUROC | F1 Score |
|---|---|---|---|---|---|
| Histogram Equalization | 0.7647 | 0.7649 | 0.7623 | 0.9427 | 0.7636 |
| Histogram Equalization followed by Gamma Correction | 0.7590 | 0.7604 | 0.7587 | 0.9396 | 0.7595 |
| Gamma Correction followed by Histogram Equalization | 0.7247 | 0.7249 | 0.7247 | 0.9349 | 0.7248 |
| Gamma Correction (1.5) | 0.7397 | 0.7416 | 0.7387 | 0.9348 | 0.7401 |
| Unsharp Masking | 0.7997 | 0.7998 | 0.7990 | 0.9536 | 0.7994 |

Table 5: Classification metrics of InceptionResNetV2 on different augmentation techniques

| Augmentation Technique | Accuracy | Precision | Recall | AUROC | F1 Score |
|---|---|---|---|---|---|
| Histogram Equalization | 0.8823 | 0.8829 | 0.8823 | 0.9631 | 0.8826 |
| Histogram Equalization followed by Gamma Correction | 0.9047 | 0.9049 | 0.9043 | 0.9711 | 0.9046 |
| Gamma Correction followed by Histogram Equalization | 0.8880 | 0.8883 | 0.8877 | 0.9642 | 0.8880 |
| Gamma Correction (1.5) | 0.8427 | 0.8429 | 0.8427 | 0.9626 | 0.8428 |
| Unsharp Masking | 0.8110 | 0.8109 | 0.8107 | 0.9451 | 0.8108 |

As evident from the aforementioned tables, it is clear that InceptionResNetV2 performs significantly better than DenseNet-201 when tasked at differentiating images that possess the traces of COVID-19 in the lung region. Even though certain confounding regions of the images were eliminated, the models are still able to classify the images with a high degree of accuracy, this insinuates that even though bias elimination was performed in a destructive fashion, there lies certain information in the lung region that can act as the centre of attention for the models.

## 6. CONCLUSIONS

The COVID-19 pandemic has impacted the lives of everyone for the past two years, and due to this, there is an urgency in developing research catering towards the common good, however, such urgency can often lead to lapses in the review process. Lately, numerous methodologies that make use of multiple publicly available datasets have emerged. However, most of these studies have failed to address the presence of bias present in the available datasets, or the bias that arises due to intermixing of datasets for the classification task. Hence, in our study, we proposed a 2-staged methodology; in stage 1, two experiments were carried out on different datasets to establish the bias present in the datasets, and thereafter, in stage 2, an image segmentation, super-resolution and CNN based pipeline along with various image augmentation techniques were proposed in order to reduce the effect of bias. Paul Mooney, Shenzhen Chest X-Ray and Montgomery County Chest X-Ray datasets were ruled out as a result of the experiments conducted in stage 1. Subsequently, following the proposed pipeline in stage 2, Chest X-Ray images augmented with Histogram Equalization followed by Gamma Correction yielded a top accuracy of 90.47% when fed into InceptionResNetV2.